\documentclass[english,preprint,superscriptaddress,showpacs,prb]{revtex4}
\usepackage[T1]{fontenc}
\usepackage[latin1]{inputenc}
\usepackage{amsmath}
\usepackage{amssymb}
\usepackage{graphicx}
\usepackage{graphicx}
\usepackage{amsfonts}
\usepackage{babel}

\input{tcilatex}

\begin{document}

\author{Alexey A. Kovalev}
\affiliation{Kavli Institute of NanoScience, Delft University of Technology, 2628 CJ
Delft, The Netherlands}
\author{Gerrit E. W. Bauer}
\affiliation{Kavli Institute of NanoScience, Delft University of Technology, 2628 CJ
Delft, The Netherlands}
\author{Arne Brataas}
\affiliation{Department of Physics, Norwegian University of Science and Technology,
N-7491 Trondheim, Norway}
\title{Magnetomechanical Torques in Small Magnetic Cantilevers}

\begin{abstract}
We study the dnamics of small magnetic cantilevers, either made from  Si covered
by a magnetic film or entirely ferromagnetic ones. The magnetomechanical
torques are found to cause line splittings in ferromagnetic resonance
spectra and magnetization reversal facilitated by mechanical degrees
of freedom. We show that the magnetomechanical torques can extend
the limits of detecting and exciting motion at the nanoscale. A "nanomotor" \ 
described here effectively transforms rf magnetic fields into mechanical
oscillations. We furthermore  propose to integrate mechanical oscillators
into magnetoelectronic devices that make use of current-induced spin-transfer torques. This
opens new possibilities for electric transducers of nanomechanical
motion.
\end{abstract}

\date{\today{}}
\maketitle

\section{Introduction}

Micro- and nanoelectromechanical systems (NEMS) \cite{Roukes:pw01,Schwab:pt05}
allow nanoscale control over as small sensors with spatial resolution on an
atomic scale \cite{Sidles:rmp95} operating at frequencies in the GHz
range.\cite{Roukes:nat03} NEMS detection of extremely small forces
corresponding to biomolecular interactions \cite{Friedsam:jpcm03} and mass
changes corresponding to single molecules \cite{Ekinci:jap04} have been
reported. Efforts have been made to merge the field of nanomechanics with that
of nanoscale magnetism. The success in the detection of a single electron spin
by magnetic resonance is a good example.\cite{Rugar:nat04} A nuclear spin
sensor that is based on imposing a coherent motion on nuclear spins has been
proposed.\cite{Bargatin:prl03} Here we focus on the possibilities provided by
the coupled motion of the strain field of a cantilever and the magnetization
of a ferromagnet.\cite{Kovalev:apl03,Kovalev:prl05} Magnetization reversal in
small magnetic clusters\cite{SpinDyn} is usually realized by external magnetic
fields \cite{Cluster,Gerrits:nat02,Back:prl98,Schumacher:prl03} or polarized
spin currents employing spin-transfer torques.
\cite{Slonc,Tsoi:prl98,Valet,Ono:DW,Wegrowe:epl99,Sun:jmmm99,Myers:sc99,Kiselev:nat03}
Along alternative mechanisms to switch the magnetization such as employing
antiferromagnets\cite{Kimel:nat04} and time-dependent magnetic
fields \cite{Sun:2005}, we suggest the effect of mechanically assisted
magnetization reversal.\cite{Kovalev:prl05} The fastest mechanical device
reported in the literature is operated with the help of magnetomotive forces
at GHz frequencies.\cite{Roukes:nat03} We propose to actuate such systems by
demagnetizing currents that provide coupling between the mechanical and the
magnetization motion most efficiently in the GHz range. Coupling of the
magnetization motion to an electric circuit via spin-transfer torques and
magnetoresistance then opens the possibility of high frequency transducers of
nanomechanical motion. In this paper, we report consequences for the
magnetovibrational coupling such as splitting of the ferromagnetic resonance
(FMR) spectrum and magnetization reversal facilitated by the mechanical
degrees of freedom. We argue that a predominantly ferromagnetic cantilever
provides stronger coupling and, when integrated into a magnetoelectronic
circuit, offers new functionalities for both mechanical and magnetic devices.
Some preliminary results have been reported already.
\cite{Kovalev:apl03,Kovalev:prl05}

The paper is organized as follows. In Sec. II, we describe our system, the
magnetic cantilever, and derive a set of equations describing
magnetomechanical motion. In Sec. III, we solve these equations in the limit
of small magnetization oscillations, finding a splitting of FMR spectra for
the resonantly coupled mode. In Sec. IV, we propose a current-induced magnetic
resonance technique in spin valves and demonstrate that magnetovibrational
modes can be detected by this method. Finally, in Sec. V, we analyze large
magnetization oscillations in the presence of magnetovibrational couplings and
demonstrate magnetization reversal assisted by the mechanical degrees of freedom.

\section{System}

We consider first a small dielectric cantilever with a single-domain
ferromagnetic layer deposited on its far end (see Fig. \ref{fig1}). A constant
external field $\mathbf{H}_{0}$ is applied along the $z$ axis. In section $3,$
we also include an oscillating field $H_{\text{y}}$ along the $y$ axis to pump
and probe the system. The effective field $\mathbf{H}_{\mathrm{eff}}$ felt by
the magnetization consists of $\mathbf{H}_{0}$ as well as the crystal
anisotropy and the demagnetizing fields. The strains are localized in the
mechanical link of the cantilever between the ferromagnetic film at one end
and the other end that is fixed. The lattice of the ferromagnet then
oscillates without internal mechanical strains, but crystalline and form
anisotropies couple the magnetic order parameter to the torsional mode of the cantilever.

Our setup consists of two weakly interacting subsystems - the magnetic and the
mechanical. It is then useful to first study the two subsystems separately.

\subsection{Magnetization motion in a single-domain ferromagnet}

The magnetization $\mathbf{M}$ of the ferromagnet precesses around an
effective magnetic field $\mathbf{H}_{\mathrm{eff}}$ according to the
Landau-Lifshitz-Gilbert equation:\cite{Gilbert:pr55}%
\begin{equation}
\frac{d\mathbf{M}}{dt}=-\gamma\mathbf{M}\times\mathbf{H}_{\mathrm{eff}}%
+\frac{\alpha}{M_{\text{s}}}\mathbf{M}\times\frac{d\mathbf{M}}{dt}%
,\label{LL+current}%
\end{equation}
where $\gamma$ denotes the gyromagnetic ratio. The phenomenological Gilbert
constant is typically $\alpha\lesssim0.01$ for metallic and $\alpha
\lesssim0.00001$ for insulating ferromagnets. The effective field is given by the functional
derivative of the free energy of the system, that is of the form (in the
lowest order in magnetizations in some specially chosen reference frame)%
\begin{equation}
E_{mg}=\frac{1}{2}D_{x}M_{x}^{2}+\frac{1}{2}D_{y}M_{y}^{2}+\frac{1}{2}%
D_{z}M_{z}^{2}-\mathbf{M}\mathbf{H}_{0},\label{Menergy}%
\end{equation}
where $D_{x}$, $D_{y}$ and $D_{z}$ describe the anisotropy of the
magnetization along the Cartesian axes $x$, $y$ and $z$, including
demagnetizing effects and crystalline anisotropy. The associated effective
field is given by $\mathbf{H}_{\mathrm{eff}}=\mathbf{H}_{0}-D_{x}%
M_{x}\mathbf{x}-D_{y}M_{y}\mathbf{y}-D_{z}M_{z}\mathbf{z}$.

\subsection{Small magnetization oscillations and dependence of FMR broadening
on shape and crystal anisotropies}

To first order, the deviations from the equilibrium magnetization in the
$z$-direction lie in the $x-y$ plane: $\mathbf{M}=m_{\text{x}}\mathbf{x}%
+m_{\text{y}}\mathbf{z}+M_{\text{s}}\mathbf{z}$, where $M_{\text{s}}$ is the
saturation magnetic moment. The magnetic susceptibilities $\chi_{\text{yy}%
}(\omega)=\left(  m_{\text{y}}/H_{\text{y}}\right)  _{\omega}$ and
$\chi_{\text{yx}}(\omega)=\left(  m_{\text{y}}/H_{\text{x}}\right)  _{\omega}$
describe the linear response of the magnetization $m_{\text{y}}$ to a (weak)
rf magnetic field $H_{\text{y}}(H_{\text{x}})$ at frequency $\omega$ and can
be found after linearizing the LLG equation in frequency space:%
\begin{equation}
\chi_{\text{yy}}(\omega)=\frac{\gamma^{2}M_{\text{s}}(H_{0}+(D_{x}%
-D_{z})M_{\text{s}})}{\omega^{2}-(1+\alpha^{2})\omega_{\text{m}}^{2}%
+i\alpha\omega(2H_{0}+(D_{x}+D_{y}-2D_{z})M_{\text{s}})}, \label{damping}%
\end{equation}
\begin{equation}
\chi_{\text{yx}}(\omega)=\frac{i\omega\gamma M_{\text{s}}}{\omega
^{2}-(1+\alpha^{2})\omega_{\text{m}}^{2}+i\alpha\omega(2H_{0}+(D_{x}%
+D_{y}-2D_{z})M_{\text{s}})}, \label{damping1}%
\end{equation}
where $\omega_{\text{m}}^{2}=(H_{0}+(D_{x}-D_{z})M_{\text{s}})(H_{0}%
+(D_{y}-D_{z})M_{\text{s}})$ is the FMR resonant frequency. Note that the FMR
damping has to be renormalized in the presence of anisotropies. The broadening
of the FMR lineshape is%
\[
\alpha^{\prime}=\alpha\frac{(H_{0}/M_{\text{s}}+(D_{x}+D_{y})/2-D_{z})}%
{\sqrt{(H_{0}/M_{\text{s}}+D_{x}-D_{z})(H_{0}/M_{\text{s}}+D_{y}-D_{z})}}%
\]

Let us consider an ellipsoid with the semi-axes $a$ and $b=c$. For small
crystalline anisotropies the anisotropy factors are defined by shape
anisotropies%
\begin{equation}%
\begin{array}
[c]{c}%
D_{x}=\frac{4\pi}{m^{2}-1}\left[  \frac{m}{2\sqrt{m^{2}-1}}\ln(\frac
{m+\sqrt{m^{2}-1}}{m-\sqrt{m^{2}-1}})-1\right] \overset{m \ll 1}{\approx}4 \pi-2\pi m ,\\
D_{y}=D_{z}=\frac{2\pi m}{m^{2}-1}\left[  m-\frac{1}{2\sqrt{m^{2}-1}}\ln
(\frac{m+\sqrt{m^{2}-1}}{m-\sqrt{m^{2}-1}})\right] \overset{m \ll 1}{\approx} \pi m ,
\end{array}
\label{an}%
\end{equation}
where $m=a/c$. The dependence of FMR broadening on the aspect ratio $m$ is
plotted in Fig. \ref{fig2}.

\subsection{Nonlinear magnetization oscillations}

Without Gilbert damping the dynamics of the magnetization can be analyzed
analytically for different magnetic anisotropies \cite{Serpico:jap03}. We
present here solutions for the {}\textquotedblleft easy
plane\textquotedblright\ anisotropy when $D_{y}=D_{z}=0$ and $\mathbf{H_{0}} =
H_{0} \mathbf{z}$
 which is relevant for
Section 4. Without the Gilbert damping but including the anisotropy factor
$D_{x}$ the LLG equations read%
\begin{equation}
\left\{
\begin{array}
[c]{c}%
\frac{dM_{\mathrm{x}}}{dt}=-\gamma M_{\mathrm{y}}H_{0}\\
\frac{dM_{\mathrm{y}}}{dt}=\gamma M_{\mathrm{x}}H_{0}+\gamma D_{x}%
M_{\mathrm{z}}M_{\mathrm{x}}\\
\frac{dM_{\mathrm{z}}}{dt}=\gamma D_{x}M_{\mathrm{y}}M_{\mathrm{x}}%
\end{array}
\right.  \label{exact}%
\end{equation}
The $x-$component of the magnetization then obeys%
\begin{equation}
\ddot{M_{\mathrm{x}}}=-(H_{0}^{2}-D_{x} E_{mg})\gamma^{2}M_{\mathrm{x}}-\frac
{\gamma^{2}}{2}D_{x}^{2} M_{\mathrm{x}}^{3}\label{duffing}%
\end{equation}
where the energy $E_{mg}=-H_{0}M_{\mathrm{z}}+D_{x}M_{\mathrm{x}}^{2}/2$ is constant when there is no Gilbert damping.

This equation describes a so-called \textquotedblleft Duffing
oscillator\textquotedblright, one of the rare examples of a non-linear dynamic
systems that can be solved analytically. The potential energy of the
oscillator has two minima. As a result, one can observe effects like periodic
motion centered at one of those minima that with increasing energy suddenly
doubles its period. When a periodic external force is applied to such an
oscillator, the system may carry out jumps between minima leading to
stochastic motion in time. Duffing's equation%
\begin{equation}
\ddot{x}-kx+2\lambda x^{3}=0 \label{duffing1}%
\end{equation}
can be integrated at once to give the first integral:%
\begin{equation}
\dot{x}^{2}-kx^{2}+\lambda x^{4}=Z \, , \label{integral}%
\end{equation}
where $Z$ is the energy of the Duffing oscillator.
By comparing Eqs. (\ref{exact},\ref{duffing}) to Eqs. (\ref{duffing1},\ref{integral}) we obtain that
$Z=\gamma^{2}(M^{2}H_{0}^{2}-E_{mg}^{2})$, $k=\gamma^{2}(D_{x} E_{mg}-H_{0}^{2})$ and 
$\lambda=\gamma^{2}D_{x}^{2}/4$. The energy minima of the
Duffing oscillator therefore correspond to the energy maxima in our system. One can distinguish three different types of solutions.

1. $Z<0$. \textquotedblleft Small amplitude\textquotedblright\ solutions when
motion is at energies close to an energy minimum (in Fig. \ref{fig3} it corresponds to two small circles and $E_{mg}>1$, the magnetization oscillates close to the perpendicular to the film direction). We express the solutions in
terms of Jacobi elliptic functions $\mathrm{dn}$:%
\begin{equation}
x=\sqrt{\frac{k}{\lambda(2-v^{2})}}\mathrm{dn}\left(  \sqrt{\frac{k}%
{\lambda(2-v^{2})}}t,v\right)  \label{case1}%
\end{equation}
where $v$ can be found from $Z=-\frac{k^{2}(1-v^{2})}{4\lambda(1-v^{2}/2)^{2}%
}.$

2. $Z=0$ corresponds to the separatrix (in Fig. \ref{fig3} it is the longest possible trajectory in a shape of a bent "8"). Here the Jacobi functions degenerate
to hyperbolic function $\mathrm{ch}$:%
\begin{equation}
x=\pm\sqrt{\frac{k}{\lambda}}\frac{1}{\mathrm{ch}(kt)} \label{case2}%
\end{equation}

3. $Z>0$ (in Fig. \ref{fig3} it corresponds to bent elliptical trajectories and $E_{mg}<1$, the magnetization trajectories are squeezed in the perpendicular to the film direction). \textquotedblleft Large amplitude\textquotedblright\ solutions via
the Jacobi functions $\mathrm{cn}$:
\begin{equation}
x=\sqrt{\frac{k}{\lambda(2v^{2}-1)}}v\:\mathrm{cn}\left(  \sqrt{\frac
{k}{\lambda(2v^{2}-1)}}t,v\right)  \label{case3}%
\end{equation}
where $Z=-\frac{k^{2}v^{2}(1-v^{2})}{\lambda(2v^{2}-1)^{2}}$. 

The analytical solution in our case can be written for the entire parameter
space with the exception of the separatrix by a single formula as follows:%
\begin{equation}%
\begin{array}
[c]{c}%
M_{x}=\sqrt{-2p_{-}}dn\left(  t\gamma\sqrt{-p_{-}/2},1-p_{+}/p_{-}\right)\overset{p_{-}\ll p_{+}}{\approx} \sqrt{-2p_{-}}\cos(t\sqrt{p_{+}/2})
\end{array}
\end{equation}
\begin{equation}%
\begin{array}
[c]{c}%
M_{y}=-\frac{\dot{M_{x}}}{\gamma H_{0}},\; M_{z}=\frac{D_{x} M_{x}^{2}-2E_{mg}}{2H_{0}}%
\end{array}
,
\end{equation}
where $p_{\pm}=H_{0}^{2}-D_{x}E_{mg} \pm H_{0}\sqrt{H_{0}^{2}-2D_{x}E_{mg}+D_{x}^{2}%
M_{s}^{2}}$. When $D_{x}>0$ this leads to the trajectories and oscillation periods
depicted in Fig. \ref{fig3}. The motion is periodic with time periods that can
be expressed by elliptic integrals $K$ as%
\begin{equation}%
\begin{array}
[c]{cc}%
T_{1}=4\sqrt{2}K\left(  p_{-}/p_{+}\right)  /\left(  \gamma\sqrt{p_{+}%
}\right)\overset{p_{-}\ll p_{+}}{\approx} 2 \pi \sqrt{2/p_{+}}  ; & E_{mg}<MH_{0}\\
T_{2}=2\sqrt{2}K\left(  1-p_{+}/p_{-}\right)  /\left(  \gamma\sqrt{-p_{-}%
}\right)\overset{|p_{-}-p_{+}|\ll |p_{-}|}{\approx} \pi \sqrt{-2/p_{-}}  . & E_{mg}>MH_{0}%
\end{array}
\label{T1/2}%
\end{equation}
The variation of the periodicity has important consequence for the
coupling to the lattice in the regime of large magnetization oscillations as
explained below.

\subsection{Cantilever oscillations and coupled magneto-mechanical equations}

Throughout this paper, we assume that the torsional oscillations of the
cantilever are small. The torsional motion of the part of the cantilever that
is not covered by the ferromagnet can be found by applying the variational
principle to the total elastic energy:\cite{Landau:59}%
\begin{equation}
E_{el}=\frac{1}{2}\int_{0}^{L}C\tau^{2}dy, \label{Eenergy}%
\end{equation}
where $\tau=\partial\varphi/dy\,\,$and $C$ is an elastic constant defined by
the shape and material of the cantilever ($C=\frac{1}{3}\mu da^{3}$ for a
plate with thickness $a$ much smaller than width $d$, $a\ll d$, $\mu$ is the
Lam$\acute{\mathrm{e}}$ constant). $T=C\tau\left(  y\right)  $ is the torque
flowing through the cantilever at point $y$. The integration is taken from the
clamping point $y=0$ until the cantilever endpoint $y=L$. The equation of
motion reads%
\begin{equation}
C\frac{\partial^{2}\varphi}{\partial y^{2}}=\rho I\frac{\partial^{2}\varphi
}{\partial t^{2}}+2\beta\rho I\frac{\partial\varphi}{\partial t},
\label{Mdynamics}%
\end{equation}
where $I=\int(z^{2}+x^{2})dzdx\simeq ad^{3}/12$ is the moment of inertia of
the cross-section about its center of mass, $\rho$ the mass density, and
$\beta$ is a phenomenological damping constant related to the quality factor
$Q$ at the resonance frequency $\omega_{\text{e}}$ as $Q=\omega_{\text{e}%
}/(2\beta)$ (at 1 GHz $Q\sim500$). \cite{Roukes:nat03} Note that $\omega_{\text{e}}$ can be
also a higher harmonic resonance frequency in what follows. The oscillating
solution has the form $\varphi=\sin(ky)(A_{1}\sin(\omega t)+A_{2}\cos(\omega
t))$, where $k=(\omega + i \beta)/c$ is the wave number, $c=c_{\text{t}}2h/d=\sqrt{C/(\rho
I)}$ and $c_{\text{t}}=\sqrt{\mu/\rho}$ is the transverse velocity of sound.
The free constants $A_{1}$ and $A_{2}$ depend on the initial conditions. The
boundary condition $\varphi|_{y=0}=0$ at the clamping point is already
fulfilled, and the boundary condition at the end $y=L$ is discussed in the following.

Combining Eqs. (\ref{Menergy},\ref{Eenergy}) and taking into account the
smallness of the magnet $\Delta L\ll L$ we can write the free energy of the
cantilever coupled to the magnetization:%
\begin{equation}
F=V(-\mathbf{M}\mathbf{H}_{0}+\frac{D_{x}}{2}\left[  M_{\mathrm{x}%
}+M_{\mathrm{z}}\varphi(L)\right]  ^{2}+\frac{D_{z}}{2}\left[  M_{\mathrm{z}%
}-M_{\mathrm{x}}\varphi(L)\right]  ^{2}+\frac{D_{y}M_{\mathrm{y}}^{2}}%
{2})+\frac{C}{2}\int_{0}^{L}\tau^{2}dy. \label{Free Energy}%
\end{equation}
Eq. (\ref{Free Energy}) demonstrates that magneto-mechanical coupling is only
possible when the factors $D_{x}$ or $D_{z}$ are non-zero (the anisotropy factor $D_{y}$ does not contribute to the 
coupling since our mechanical motion is rotationally invariant with respect to the axis $y$ and one needs anisotropies 
$D_{x}$ and $D_{z}$ to break the invarance). For
small $\varphi=\varphi(L)$, where $\varphi(y)$ is the torsion angle at
position $y$ of the cantilever, the effective field $\mathbf{H}_{\mathrm{eff}%
}=-\frac{\partial F}{\partial\mathbf{M}}$ is
\[
\mathbf{H}_{\mathrm{eff}}=(D_{x}M_{\mathrm{z}}\varphi-D_{x}M_{\mathrm{x}%
})\mathbf{x}+D_{x}M_{\mathrm{x}}\varphi\mathbf{z}-(D_{z}M_{\mathrm{x}}%
\varphi+D_{z}M_{z})\mathbf{z}-D_{z}M_{\mathrm{z}}\varphi\mathbf{x}-D_{y}%
M_{y}+\mathbf{H_{0}}%
\]
where $D_{x}$, $D_{y}$ and $D_{z}$ describe the anisotropies ($D_{x}\simeq
4\pi$ for a thin film without crystal anisotropies). The equation of motion of
the magnetization $\mathbf{M}$ in the presence of the mechanical degree of
freedom then reads%
\begin{equation}
\frac{d\mathbf{M}}{dt}=-\gamma\mathbf{M}\times\mathbf{H}_{\mathrm{eff}}%
+\frac{\alpha}{M_{\text{s}}}\mathbf{M}\times\left(  \frac{d\mathbf{M}}%
{dt}\right)  _{\text{cant}}, \label{LLG}%
\end{equation}
where the derivative $\left(  \frac{d\mathbf{M}}{dt}\right)  _{\text{cant}%
}=\frac{d\mathbf{M}}{dt}+\frac{d\mathbf{\varphi}}{dt}(-M_{\mathrm{z}%
}\mathbf{x}+M_{\mathrm{x}}\mathbf{z})$ is taken in the reference system of the
cantilever since the magnetization damping is caused by interactions of the
magnetization with the bulk of the cantilever.

By applying the variational principle to Eq. (\ref{Free Energy}) we obtain Eq.
(\ref{Mdynamics}) and the second boundary condition for its solutions. The
magnetovibrational coupling can then be treated as a boundary condition to the
mechanical problem, which is expressed as the torque $C\tau|_{y=L}$ exerted by
the magnetization on the edge of the cantilever:%
\begin{equation}
C\tau|_{y=L}=\frac{1}{\gamma}\left(  \frac{d\mathbf{M}}{dt}+\gamma
\mathbf{M}\times\mathbf{H}_{0}\right)  |_{y}, \label{bound}%
\end{equation}

This boundary condition is equivalent to the conservation law of the
mechanical angular momentum written for the tip of the cantilever. Let us
introduce an angular momentum $\mathbf{V}^{\mathrm{el}}(y)$ for a thin slice
at point $y\in\left\{  0,L\right\}  $ (without magnetic overlayer), then the
conservation law is $d\mathbf{V}^{\mathrm{el}}\left(  y\right)  /dt=\mathbf{T}%
\left(  y\right)  ,\,\,$where $\mathbf{T}(y)$ is the torque flowing into the
slice. This equation is modified by the coupling to the magnet in a region
$y\in\left\{  L,L+\Delta L\right\}  $ ($\Delta L$ is the length of the
cantilever covered by the magnetic layer) as:%
\begin{equation}
\frac{d}{dt}\left(  \mathbf{V}^{\mathrm{el}}\left(  L\right)  +(-\frac
{1}{\gamma})\mathbf{M}\left(  L\right)  V\right)  =\mathbf{T}\left(  L\right)
+\mathbf{T}_{\text{field}}. \label{motion}%
\end{equation}
where $\mathbf{T}_{\text{field}}=V\mathbf{M}\times\mathbf{H}_{0}$, $V$ is the
volume of the magnet and $\mathbf{T}(y)|_{y}=-C\tau\left(  y\right)  $. When
$\Delta L\ll L,$ internal strains in the magnetic section may be disregarded
and the magnetization torque can be treated as a boundary condition Eq.
(\ref{bound}).

The coupling of Eqs. (\ref{LL+current}) and (\ref{Mdynamics}) can be made
explicit:%
\begin{equation}%
\begin{array}
[c]{c}%
\partial\varphi/dy|_{y=L}=\frac{1}{C\gamma}\left(  \frac{dM_{\text{y}}}%
{dt}+\gamma\mathbf{M}\times\mathbf{H}_{0}\right) \\
\mathbf{H}_{\mathrm{eff}}=(D_{x}M_{\mathrm{z}}\varphi-D_{x}M_{\mathrm{x}%
})\mathbf{x}+D_{x}M_{\mathrm{x}}\varphi\mathbf{z}-(D_{z}M_{\mathrm{x}}%
\varphi+D_{z}M_{z})\mathbf{z}-D_{z}M_{\mathrm{z}}\varphi\mathbf{x}-D_{y}%
M_{y}+\mathbf{H_{0}}%
\end{array}
\label{MME}%
\end{equation}

\subsection{Entirely ferromagnetic cantilever}

It is straightforward to generalize the free energy Eq. (\ref{Free Energy}) to
the case when the whole cantilever is covered by a ferromagnetic film or when
the whole cantilever is ferromagnetic. The free energy then has the form:%
\begin{equation}
F=V(-\mathbf{M}\mathbf{H}_{0}+\frac{D_{x}}{2}\left[  M_{\mathrm{x}%
}+M_{\mathrm{z}}\bar{\varphi}\right]  ^{2}+\frac{D_{z}}{2}\left[
M_{\mathrm{z}}-M_{\mathrm{x}}\bar{\varphi}\right]  ^{2}+\frac{D_{y}%
M_{\mathrm{y}}^{2}}{2})+\frac{C}{2}\int_{0}^{L}\tau^{2}dy \label{Free Energy1}%
\end{equation}
where $\bar{\varphi}=\frac{1}{L}\int_{0}^{L}\varphi(y)dy$ is the average angle
of torsion. In this case the system of the magnetomechanical equations
becomes:%
\begin{equation}%
\begin{array}
[c]{c}%
\frac{d\mathbf{M}}{dt}=-\gamma\mathbf{M}\times\mathbf{H}_{\mathrm{eff}}%
+\frac{\alpha}{M_{\text{s}}}\mathbf{M}\times\left(  \frac{d\mathbf{M}}%
{dt}\right)  _{\text{cant}}\\
C\frac{\partial^{2}\varphi}{\partial y^{2}}=\rho I\frac{\partial^{2}\varphi
}{\partial t^{2}}+\frac{S}{\gamma}\left(  \frac{dM_{\text{y}}}{dt}%
+\gamma\mathbf{M}\times\mathbf{H}_{0}\right)  +2\beta\rho I\frac
{\partial\varphi}{\partial t}\\
C\partial\varphi/dx|_{x=L}=0\\
\mathbf{H}_{\mathrm{eff}}=(D_{x}M_{\mathrm{z}}\bar{\varphi}-D_{x}%
M_{\mathrm{x}})\mathbf{x}+D_{x}M_{\mathrm{x}}\bar{\varphi}\mathbf{z}%
-(D_{z}M_{\mathrm{x}}\bar{\varphi}+D_{z}M_{z})\mathbf{z}-D_{z}M_{\mathrm{z}%
}\bar{\varphi}\mathbf{x}-D_{y}M_{y}+\mathbf{H_{0}}%
\end{array}
\label{covered}%
\end{equation}
The coupling can be much more efficient in the latter as compared to the
former system since the volume of the magnet is larger. The disadvantage is
that the form anisotropy wants to align the magnetization along the cantilever
making the subsystems uncoupled. Another disadvantage, the possibility of spin
wave excitations, is discussed in the next section.

\section{Small magnetization oscillations}

\subsection{Magnet at the tip of cantilever}

To first order, the deviations from the equilibrium magnetization in the
$z$-direction lie in the $x-y$ plane: $\mathbf{M}=m_{\text{x}}\mathbf{x}%
+m_{\text{y}}\mathbf{y}+M_{\text{s}}\mathbf{z}$, where $M_{\text{s}}$ is the
saturation magnetic moment (see Fig. 1). For small $\varphi$, the effective
field oscillates in the $x-y$ plane: $\mathbf{H}_{\mathrm{eff}}=(D_{x}%
M_{\mathrm{\text{s}}}\varphi-D_{x}M_{\mathrm{x}})\mathbf{x}+D_{x}%
M_{\mathrm{x}}\varphi\mathbf{z}-(D_{z}M_{\mathrm{x}}\varphi+D_{z}%
M_{\mathrm{\text{s}}})\mathbf{z}-D_{z}M_{\mathrm{\text{s}}}\varphi
\mathbf{x}-D_{y}M_{y}+\mathbf{H_{0}}$, ($D_{x}\simeq4\pi$ for a thin film
without crystal anisotropy). Note that the coupling does not rely on a strong
magnetocrystalline field, since the surface forces of demagnetizing currents
also provide a restoring torque. For a thin ferromagnetic layer the latter
dominates and the magnetization does not precess, but oscillates like a
pendulum in the $y-z$ plane due to the oscillating field in the $x$-direction
$\mathbf{H}_{\mathrm{eff}}=D_{x}M_{\text{s}}\varphi\mathbf{x}$.

By using Eq. (\ref{bound}) with the substitution of $\varphi$ in $\tau$,
$C\tau|_{y=L}=Ck\cos(kL)(A_{1}\sin(\omega t)+A_{2}\cos(\omega t))=Ck\varphi
\cot(kL)$, we obtain in frequency space and to first order in the
magnetization oscillations:%
\begin{align}
Ck\varphi\cot(kL)  &  =\frac{V}{\gamma}(-i\omega m_{\text{y}}-\gamma
H_{0}m_{\text{x}})\label{Fourier}\\
&  \approx-C\varphi\frac{L}{2c^{2}}(\omega^{2}+2i\beta\omega-\omega_{\text{e}%
}^{2}), \label{Fourierexp}%
\end{align}
where in the second line the $\cot$ has been expanded close to the resonance
frequency $\omega_{\text{e}}=c\pi(1/2+s)/L$ ( $s$ is integer, here we
concentrate mainly on coupling to the first harmonic and $s=0$).

The magnetic susceptibilities $\chi_{\text{yy}}(\omega)=\left(  m_{\text{y}%
}/H_{\text{y}}\right)  _{\omega}$ and $\chi_{\text{yx}}(\omega)=\left(
m_{\text{y}}/H_{\text{x}}\right)  _{\omega}$ describe the linear response of
the magnetization $m_{\text{y}}$ to a (weak) rf magnetic field $H_{\text{y}%
}(H_{\text{x}})$ at frequency $\omega$. Generalization of Eqs (\ref{damping}%
,\ref{damping1}) in the presence of magneto-vibrational coupling can be found
after writing LLG equation in frequency space with use of Eq. (\ref{Fourier}),
and taking into account smallness of the Gilbert damping $\alpha$:
\begin{equation}
\chi_{\text{yy}}(\omega)=\frac{\gamma^{2}M_{\text{s}}^{2}(D_{x}-D_{z}%
+H_{0}(1-g\tan(kL))/M_{\text{s}})}{\omega^{2}-\omega_{\text{m}}^{2}%
+2i\alpha^{\prime}\omega\omega_{\text{m}}+\left[  \omega^{2}-H_{0}(H_{0}%
+(D_{y}-D_{z})M_{\text{s}})\right]  g\tan(kL)} \label{suscept}%
\end{equation}
\begin{equation}
\chi_{\text{yx}}(\omega)=\frac{i\omega\gamma M_{\text{s}}}{\omega^{2}%
-\omega_{\text{m}}^{2}+2i\alpha^{\prime}\omega\omega_{\text{m}}+\left[
\omega^{2}-H_{0}(H_{0}+(D_{y}-D_{z})M_{\text{s}})\right]  g\tan(kL)}
\label{suscept1}%
\end{equation}
where $g=M_{\text{s}}^{2}V(D_{x}-D_{z})c^{2}/\left(  CL\omega
_{\text{e}}^{2}\right)  =(2/\pi)^{2}(D_{x}-D_{z})(L/a)^{2}(V/V_{c})(M_{\text{s}}^{2}/\mu)$ is the magnetovibrational coupling constant ($V$
and $V_{c}$ are the volumes of the magnet and the cantilever respectively) and
the unperturbed resonance frequency $\omega_{\text{m}}^{2}=(H_{0}+(D_{x}%
-D_{z})M_{\text{s}})(H_{0}+(D_{y}-D_{z})M_{\text{s}})$. Note that the coupling
constant is defined by the material parameters in the term 
$M_{\text{s}}^{2}/\mu$, by the geometry in the term $(L/a)^{2}(V/V_{c})$
and by both the geometry and the material parameters in the term $(D_{x}%
-D_{z})$. The imaginary
part of $\chi_{\text{yy}}(\omega)$ is proportional to the FMR\ absorption signal.

When the cantilever with the magnet is subjected to the rf
magnetic field, the magnetomechanical torques induce a mechanical motion. In
effect we have then constructed a nano-scale motor that transforms the
magnetic field oscillations into mechanical motion with amplitudes that are
given by the susceptibility $\chi_{\varphi\text{y}}(\omega)=\left(
\varphi/H_{\text{y}}\right)  _{\omega}.$ The latter follows from the LLG
equation in frequency space using Eq. (\ref{Fourier}), and taking into account
the smallness of the Gilbert damping $\alpha$:
\begin{equation}
\chi_{\varphi\text{y}}(\omega)=\frac{i\omega g\tan(kL)}{\omega^{2}%
-\omega_{\mathrm{m}}^{2}+2i\alpha^{\prime}\omega\omega_{\text{m}}+\left[
\omega^{2}-H_{0}(H_{0}+(D_{y}-D_{z})M_{\text{s}})\right]  g\tan(kL)}
\label{Nmotor}%
\end{equation}

In the absence of the external field $H_{0}$ the resonance frequencies in the
vicinity of the resonance frequency $\omega_{\text{e}}$ can be found after
expanding $\cot(kL)$ as in Eq. (\ref{Fourierexp}):%
\begin{equation}
\omega_{1(2)}=\sqrt{\frac{1}{2}}\left[  \omega_{\text{e}}^{2}+\omega
_{\mathrm{m}}^{2}+g\omega_{\text{e}}^{2}\pm\left(  ((\omega_{\text{e}}%
+\omega_{\mathrm{m}})^{2}+g\omega_{\text{e}}^{2})((\omega_{\text{e}}%
-\omega_{\text{m}})^{2}+g\omega_{\text{e}}^{2})\right)  ^{1/2}\right]  ^{1/2}.
\label{resonance}%
\end{equation}
When the external field $H_{0}$ is smaller than the demagnetizing field, Eq.
(\ref{resonance}) holds with $\omega_{\text{m}}^{2}=(H_{0}+(D_{x}%
-D_{z})M_{\text{s}})(H_{0}+(D_{y}-D_{z})M_{\text{s}})$. This fact can be used
to tune the FMR frequency in order to match the elastic frequency. In this
limit of strong shape anisotropy (for example a thin ferromagnetic film with
$D_{x}\sim4\pi$) and when $H_{0}G/(M_{\text{s}}\beta\omega_{\text{e}}^{2}%
)\ll1,$ Eqs. (\ref{suscept},\ref{Nmotor}) simplify to%
\begin{equation}
\chi_{\text{yy}}(\omega)\approx\frac{\gamma^{2}M_{\text{s}}^{2}(D_{x}-D_{z}%
)}{\omega^{2}-\omega_{\text{m}}^{2}+2i\alpha^{\prime}\omega\omega_{\text{m}%
}+\omega^{2}g\tan(kL)}, \label{suscept2}%
\end{equation}
\begin{equation}
\chi_{\varphi\text{y}}(\omega)\approx\frac{i\omega g\tan(kL)}{\omega
^{2}-\omega_{\text{m}}^{2}+2i\alpha^{\prime}\omega\omega_{\text{m}}+\omega
^{2}g\tan(kL)}. \label{Nmotor1}%
\end{equation}
Imaginary part of $\chi_{\text{yy}}(\omega)$ corresponding to the rf
absorption is plotted in Fig. \ref{fig5}. We observe a typical anticrossing
behavior between an optically active and non-active mode, with level repulsion
and transfer of oscillator strength. The intrinsic damping of the mechanical
system (for MEMS $Q$ factors can reach $10^{4},$ which quickly deteriorate
with decreasing size, however) imposes an extra damping on the magnetization
dynamics, which close to the mode crossing may dominate the intrinsic damping
due to a small Gilbert constant $\alpha$. In case $\beta/\omega>\alpha^{\prime}$ the
damping growth when we move from purely magnetic motion into purely mechanical
motion along one of the lines in Fig. \ref{fig4}. In case $\beta/\omega
<\alpha^{\prime}$ the damping diminishes along such line.

The amplitude of mechanical oscillations along one of the spectrum lines in
Fig. \ref{fig4} is plotted in Fig. 6. The efficiency of the nanomotor is
maximal at resonance.

\subsection{Ferromagnetic cantilever}

Eq. (\ref{covered}) describes magnetomechanical dynamics for a ferromagnetic
cantilever or a dielectric cantilever covered by a ferromagnet in the whole.
The LLG equation is exactly the same as in case of previous subsections apart
from the fact that the torsion angle should be averaged over the cantilever.
Following Eq. (\ref{covered}) and for frequencies close to the mechanical
resonance, the coupling can be written in Fourier space as follows
\begin{equation}
-C\bar{\varphi}\frac{L}{2c^{2}}(\omega^{2}+2i\beta\omega-\omega_{\text{e}}%
^{2})=\frac{V_{c}}{2\gamma}(-i\omega m_{\text{y}}-\gamma H_{0}m_{\text{x}})
\label{Fcovered}%
\end{equation}
This result is identical to the expansion in Eqs. (\ref{Fourier}%
,\ref{Fourierexp}) after substituting $V_{c}/2$ by $V$. In the expression for
the coupling constant $g$ in Eqs (\ref{suscept},\ref{suscept1}) we have
$V/V_{c}=1/2$. The coupling constant is thus increased in comparison to the
case of a small magnet at the tip of the cantilever in which $V/V_{c}\ll1$.

The density of a metallic single crystal cantilever (Fe) is higher $\rho \sim%
8000\,\,\mathrm{kg/m}^{3}$ and the Lam$\acute{\mathrm{e}}$ constant $\mu \sim 100 \mathrm{GPa}$ \cite{shear}
is of the same order as for Si. Consequently, the
metallic cantilever has to be smaller in order to have
the same resonance frequency with the Si cantilever. Most importantly, we have
to fulfill the condition $H_{0}>D_{z}M_{\text{s}}$ that forces the equilibrium
magnetization direction along the $z-$axis (alignment with the $y-$axis would
result in zero coupling). This condition can be easily fulfilled in our
geometry by not too strong magnetic fields since the largest demagnetizing
factor is $D_{x}$.

It follows from Eqs. (\ref{suscept},\ref{suscept1}) that a ferromagnetic
cantilever is more suitable for observating the magneto-mechanically coupled
modes since the coupling strength is larger by the factor $V_{c}/2V$ compared
to the system with a small magnet at the tip of the cantilever. Close to the
mechanical resonance frequency both systems will behave identically.

\subsection{Observation}

The magnetovibrational coupling in our cantilever is observable by FMR and
calorimetric techniques,\cite{Moreland:rsi00} but the signal of nanoscale
magnets is small. It might therefore be preferable to detect the resonance by
the static deflection of the same cantilever due to an additional constant
magnetic field $\mathbf{H}_{T}$ along the $x$-axis, as described by Lohndorf
\textit{et al}.\cite{Lohndorf:apl00}. In our approximation the field
$\mathbf{H}_{\text{T}}$ creates a torque $\gamma M_{\text{y}}H_{\text{T}}%
\vec{x}$, whose modulation at the FMR conditions should be detectable.

Since the vibrational frequencies of state-of-the-art artificial structures
are relatively low, the use of soft ferromagnets (such as permalloy), is
advantageous. The magnetic mode frequencies are then determined by shape
anisotropies. The FMR frequency and the mechanical resonance frequency should
not differ by much more than $\Delta\omega\sim\omega\sqrt{g}$ for a pronounced
effect. A Si cantilever with $a\times d\times L=\left(  1\times5\times
50\right)  \mu m$ $\left(  C=10^{-13}Nm^{2}\right)  $ has a torsional
resonance frequency of the order of $\omega_{\text{e}}=10$ MHz. Taking our
ferromagnetic layer of dimensions $a1\times d\times\Delta L=50nm\times5\mu
m\times5\mu m$ (thickness, width and length), then $\omega\sqrt{g}\sim100\,\,
KHz$, meaning that we should tune the magnetic resonance to $\omega_{\text{e}%
}\pm100\,\, KHz$ to observe the \textquotedblleft polariton\textquotedblright.
The necessary rf field $H_{\text{x}}$ depends on the viscous dampings of
mechanical and magnetization motion. At low frequencies additional sources of
damping complicate measurements\cite{Cochran:prb89} and the coherent motion of
the magnetization can be hindered by domain formation. Coupling to higher
resonance modes\cite{Rabe:rsi96} or structuring of the ferromagnet may help to
carry out measurements.

An actual observation of the predicted splittings would give information about
\textit{e.g.} the magnetic moment of the film and the broadening would yield
the quality factor of the elastic motion. From a technological point of view
the tunable damping due to the magnetovibrational coupling might be
interesting for optimizing switching speeds. A ferromagnet effectively absorbs
microwaves and turns them into a precessing magnetization, which via the
magnetovibrational coupling can be transformed into a coherent mechanical
motion. On the other hand, the ferromagnet may interact with the mechanical
motion, to cause a magnetization precession, which in turn emits polarized
microwaves. The emission in the coupled regime is more energy efficient in
comparison with a fixed magnetic dipole emission in the case of small Gilbert
constant but relatively low mechanical quality factor. This might be
interesting for \textit{e.g}. on-chip communication applications.

In this section we have calculated the magnetic susceptibility of a system
with magnetovibrational coupling by magnetocrystalline fields or via surface
forces of demagnetizing currents. A condition for effective energy transfer
from an external rf magnetic field into mechanical motion and \textit{vice
versa} has been established. FMR spectra are predicted to split close to the
resonance, and to strongly depend on the mechanical damping. The predicted
effects should be observable with existing technology, but a further reduction
of system size would strongly enhance them.

\section{Probing magnetization dynamics by spin-transfer torques}

Microwave fields are a perfect tool for probing magnetization dynamics,
\emph{e.g.} by FMR. However, with shrinking size of the magnets such methods
become increasingly difficult since the power absorbed by the magnet becomes
very small. In this section, we propose a method to probe the magnetization
dynamics in which the role of the microwave field is taken over by 
spin-transfer torques. By driving an AC current through F$|$N$|$F spin valve
structures with fixed and free layer magnetizations that are canted with
respect to each other, we can excite the free layer motion (Fig. \ref{fig7}) (excitation of the
magnetization motion by DC currents has already been realized  \cite{Krivorotov:sc05}) 
. The magnetization dynamics
should in turn influence the resistance of the device that can be measured.
When the magnetic layer has a mechanically active extension, as in Fig.
\ref{fig7}, we can also detect the magneto-vibrational mode.

\subsection{Electrical detection of FMR}

We consider here F1$|$N$|$F2 multilayer structures with perpendicular
magnetizations $\mathbf{m}_{1}$ and $\mathbf{m}_{2}$ assuming $\mathbf{m}_{2}$
fixed by anisotropies or exchange biasing. An AC current exerts an oscillating
torque $\tau_{1}$ on the magnetization $\mathbf{m}_{1}$:\cite{Kovalev:prb02}%
\begin{equation}
\frac{\tau_{1}}{I_{0}}=\frac{\hbar}{2e}\frac{1+R_{\uparrow\downarrow}%
/R_{1}-\beta P_{1}/P_{2}}{(1+R_{\uparrow\downarrow}/R_{1})(1+R_{\uparrow
\downarrow}/R_{2})-\beta^{2}} \label{torq1}%
\end{equation}
with $\beta=\cos\theta,$ $4R_{1(2)}=1/G_{1(2)\uparrow}+1/G_{1(2)\downarrow
}-2R_{\uparrow\downarrow}$, $4R_{1(2)-}=1/G_{1(2)\uparrow}-1/G_{1(2)\downarrow
}$, $P_{1(2)}=R_{1(2)-}/R_{1(2)}$, and $2R_{\uparrow\downarrow}=1/G_{1\uparrow
\downarrow}^{r}+1/G_{2\uparrow\downarrow}^{r},$ where $G_{1(2)\uparrow}$ and
$G_{1(2)\downarrow}$ are conductances of the left (right) ferromagnet
including the left (right) normal layer, $G_{1\uparrow\downarrow}^{r}$ and
$G_{2\uparrow\downarrow}^{r}$ are mixing conductances of the middle normal
metal with adjacent ferromagnet interfaces. The resulting magnetization
dynamics causes oscillations of the resistance of the multilayer structure
according to\cite{Kovalev:prb02}%
\begin{equation}
\Re(\theta)=R+R_{1}+R_{2}-\frac{R_{\uparrow\downarrow}(R_{1-}+\beta
R_{2-})^{2}+(1-\beta^{2})(R_{1-}^{2}R_{2}+R_{2-}^{2}(R_{1}+R_{\uparrow
\downarrow}))}{(R_{\uparrow\downarrow}+R_{1})(R_{\uparrow\downarrow}%
+R_{2})-\beta^{2}R_{1}R_{2}}. \label{aMR}%
\end{equation}
In the following, we consider only small deviations of $\mathbf{m}_{1}$ from
the perpendicular to $\mathbf{m}_{2}$ direction. All the equations can then be
rewritten keeping only the leading terms with respect to small $\beta$. The
torque $\tau_{1}$ is proportional to the current $I_{0}$ in this
approximation:%
\begin{equation}
\tau_{1}=I_{0}\frac{\hbar}{2e}\frac{1+R_{\uparrow\downarrow}/R_{1}%
}{(1+R_{\uparrow\downarrow}/R_{1})(1+R_{\uparrow\downarrow}/R_{2}%
)}=V|\mathbf{M}\times\mathbf{H}_{x}| \label{torq2}%
\end{equation}
where we defined an effective rf field $\mathbf{H}_{x}$ along the axis $x$
that creates the same torque as the oscillating AC current. Thus the response
function $\left(  m_{\text{y}}/I_{0}\right)  _{\omega}=K\chi_{\text{yx}%
}(\omega)\,\,$with
\begin{equation}
KM_{\text{s}}V=\frac{\hbar}{2e}\frac{1+R_{\uparrow\downarrow}/R_{1}%
}{(1+R_{\uparrow\downarrow}/R_{1})(1+R_{\uparrow\downarrow}/R_{2})},
\label{prop}%
\end{equation}
and the susceptibility can be found from Eq. (\ref{damping1}). To the first
order in the mechanical damping constant $\beta$ we can write the dynamical
impedance as:%
\begin{equation}
Z(\omega)=R+R_{1}+R_{2}-\frac{R_{\uparrow\downarrow}(R_{1-}^{2}+2I_{0}%
K\chi_{\text{yx}}(\omega)R_{2-}R_{1-})+(R_{1-}^{2}R_{2}+R_{2-}^{2}%
(R_{1}+R_{\uparrow\downarrow}))}{(R_{\uparrow\downarrow}+R_{1})(R_{\uparrow
\downarrow}+R_{2})}. \label{aMR1}%
\end{equation}
$\left\vert Z(\omega)\right\vert $ normalized by the resistance of the locked
perpendicular magnetizations is plotted in Fig. \ref{fig8}. The parameters of
the spin valve are the same with the symmetric setup of ref. \cite{Kovalev:prb02} The current
amplitude $I_{0}$ is chosen to correspond to magnetization oscillations of
$15$ degrees which appears to be readily experimentally accessible. In Fig.
\ref{fig8} the FMR resonance corresponds to the dip in the absolute value of
the impedance that should be easily detectable.

\subsection{Electrical detection of magnetovibrational mode}

The method described in the previous chapter can also be used for detecting
the magnetovibrational modes. The ferromagnet F1 is extended forming a
ferromagnetic cantilever (Fig. \ref{fig7}). The magnetization $\mathbf{m}_{2}$
can have two directions (bold and dashed lines in Fig. \ref{fig7}) that are
equally suitable for this purpose. In principle, the direction plotted by
dashed line can cause a constant torsion of the cantilever when a constant
current is sent through the spin valve since in this case the spin-transfer
torque is transferred to the mechanical torque via the shape anisotropy. We
concentrate here on the direction of $\mathbf{m}_{2}$ indicated by the bold
arrow which can be easier realized in experiment. When an AC current is sent
through the system the oscillating torque causes oscillations of the
magnetization $\mathbf{m}_{1}$ as explained in the previous section. At
resonance, we can observe the magnetovibrational mode as a splitting of the
dip in the absolute value of the impedance plotted in Fig. \ref{fig9}. The
width of the dips is a measure of the damping that is approximately half of
the width in Fig. \ref{fig8} for chosen parameters, which is consistent with
the results of Sec 3.1 .

In a long cantilever with a finite exchange stiffness, the macrospin
magnetization motion can be complicated by spin waves. The effective field for
the LLG Eq. (\ref{LL+current}) in the presence of nonuniform magnetization
reads%
\begin{equation}
\mathbf{H}_{\mathrm{eff}}=-D_{x}M_{\mathrm{x}}\mathbf{x}-D_{y}M_{y}%
\mathbf{y}-D_{z}M_{z}\mathbf{z}+\frac{2A}{M_{\text{s}}^{2}}\bigtriangledown
^{2}\mathbf{M}+\mathbf{H_{0}} \label{spinwaves}%
\end{equation}
where $A$ is the exchange stiffness. The lowest-energy bulk spin wave mode is
along the cantilever with the frequency that can be found from the LLG
equation written in Fourier space:
\begin{equation}
\left[
\begin{array}
[c]{cc}%
-i\omega & \gamma(\frac{2A}{M_{\text{s}}}k^{2}+(D_{y}-D_{z})M_{\text{s}}%
+H_{0})\\
-\gamma(\frac{2A}{M_{\text{s}}}k^{2}+(D_{x}-D_{z})M_{\text{s}}+H_{0}) &
-i\omega
\end{array}
\right]  \left[
\begin{array}
[c]{c}%
m_{x}/M_{\text{s}}\\
m_{y}/M_{\text{s}}%
\end{array}
\right]  =0 \label{LLGspinwave}%
\end{equation}
where $\mathbf{M}=(m_{\text{x}}\mathbf{x}+m_{\text{y}}\mathbf{y})e^{i(\omega
t+\mathbf{k}\cdot\mathbf{r})}+M_{\text{s}}\mathbf{z}$. The resonance frequency
is $\omega_{sw}=\gamma\sqrt{(\frac{2A}{M_{\text{s}}}k^{2}+(D_{x}%
-D_{z})M_{\text{s}}+H_{0})(\frac{2A}{M_{\text{s}}}k^{2}+(D_{y}-D_{z}%
)M_{\text{s}}+H_{0})}$. We estimate the difference between frequencies of the
macrospin mode with $k=0$ and the longest wavelength mode with $k=\pi/L$. For
a thin film $\gamma\sqrt{4\pi M_{\text{s}}\frac{2A}{M_{\text{s}}}(\pi/L)^{2}%
}\approx0.2\text{ GHz,}$ where we adopted $A=2\times10^{-11}\mathrm{J/m}$,
$M_{\text{s}}=1.4\times10^{6}\mathrm{A/m}$ and $L=1\ \operatorname{%
\protect\rule{0.1in}{0.1in}
m}$. Since we excite the mechanical motion monochromatically by the AC currents, it is
sufficient then to have the mode splitting larger than the broadening. This shows that in principle 
we can design the magnetic and mechanical
subsystems to avoid bulk spin wave generation provided the mechanical damping
as well as the Gilbert damping are relatively small. Note that by applying an
antiferromagnetic layer on top of the cantilever we can strengthen exchange
stiffness thus diminishing the possibility of spin waves even further.

It is also possible to include coupling of the mechanical motion to spin waves
but this regime will not be considered here.

\section{Large magnetization cones and magnetization reversal in the presence
of coupling}

In this section, we extend the linearized magnetization motion to large angles
relevant \emph{e.g.} for the magnetization reversal.

\subsection{Resonant magnetization oscillations and reversal}

We consider resonant oscillations of the mechanical and
magnetic degrees of freedom. We restrict ourself here to the case when only
one anisotropy direction is present (\emph{e.g.} $D_{y}=D_{z}=0$) which is
relevant for very thin ferromagnetic films with and easy plane anisotropy when
$D_{x}\sim4\pi$.

The coupling of Eqs. (\ref{LL+current}) and (\ref{Mdynamics}) is%
\begin{equation}%
\begin{array}
[c]{c}%
\frac{\partial\varphi}{\partial y}|_{y=L}=\frac{1}{C\gamma}\left(
\frac{dM_{\text{y}}}{dt}+\gamma\mathbf{M}\times\mathbf{H}_{0}|_{y}\right)
\,,\\
\mathbf{H}_{\mathrm{eff}}=(D_{x}M_{\mathrm{z}}\varphi-D_{x}M_{\mathrm{x}%
})\mathbf{x}+D_{x}M_{\mathrm{x}}\varphi\mathbf{z}+\mathbf{H}_{0}\,,
\end{array}
\label{MME1}%
\end{equation}

We first address the maximal possible coupling strength of a system described
by Eq. (\ref{MME1}). Consider the two subsystems oscillating at a common
frequency $\omega$. The total mechanical energy is then $E_{me}=\rho
IL\omega^{2}\varphi_{0}^{2},$ where $\varphi_{0}$ is the maximal angle of the
torsional motion. By equipartition this energy should be of the order of the
magnetic energy $E_{mg}=M_{\text{s}}VH_{0}$. The maximal angle would
correspond to the mechanical motion induced by full transfer of the magnetic
energy to the lattice. By equalizing those energies, we find an estimate for
the maximal angle of torsion $\varphi_{0}=\sqrt{M_{\text{s}}VH_{0}/\left(
\rho IL\omega^{2}\right)  }$. The coupling between the subsystems can be
measured by the distribution of an applied external torque (\textit{e.g}.
applied by a magnetic field) over the two subsystems. The total angular
momentum flow into the magnetic subsystem by the effective magnetic field is
$(M_{\text{s}}V/\gamma)\omega,$ whereas that corresponding to the mechanical
subsystem at the same frequency is $\left(  \rho IL\omega\varphi_{0}\right)
\omega.$ Their ratio is $\varphi_{0}\omega/(\gamma H_{0})$. The maximum angle
$\varphi_{0}$ derived above is therefore also a measure of the coupling
between the magnetic and mechanical subsystems. This estimate is consistent
with the coupling strength of polariton modes at resonance of $g=\varphi
_{0}^{2}\omega^{2}/(\gamma H_{0})^{2}\approx\varphi_{0}^{2}D_{x}M_{\text{s}%
}/H_{0}$ in Eqs. (\ref{suscept},\ref{suscept1}) (in this estimate we consider
the case of not too strong external fields when $\omega\approx\gamma
\sqrt{H_{0}D_{x}M_{\text{s}}}$). An estimate for a cantilever with
$\rho=2330\,\,\mathrm{kg/m}^{3}$ (Si) and $d=100$ $\mathrm{nm}$ ($\omega\sim$1
GHz) leads to $g=\varphi_{0}^{2}D_{x}M_{\text{s}}/H_{0}\sim(L/a)^{2}%
(M_{\text{s}}^{2}/\mu)\sim10^{-3}$. Increasing $(L/a)$ and
$M_{\text{s}}$ or decreasing Lam$\acute{\mathrm{e}}$ constant $\mu$ is
beneficial for the coupling.

Magnetization reversal by a magnetic field in the coupling regime can be
realized even without any damping by transferring magnetic energy into the
mechanical system. Since we find that $\varphi_{0}\ll1$ for realistic
parameters, the subsystems undergo many precessions/oscillations before the
switching is completed. The switching is then associated with a slow time
scale corresponding to the global motion governed by the coupling or a
weak damping relative to a fast time scale characterized by the
Larmor frequency. The equation of motion for the slow dynamics (the envelope
functions) can be derived by averaging over the rapid oscillations. To this
end, we substitute LLG Eq. (\ref{LLG}) and first of Eqs. (\ref{MME1}),
linearized in the small parameters $\alpha$, $\beta$ and $\varphi_{0}$, into
the equations for the mechanical and the magnetic energies:%
\begin{equation}%
\begin{array}
[c]{c}%
\frac{d}{dt}E_{me}=-2\beta E_{me}+C\tau|_{x=L}\frac{d\varphi}{dt}|_{x=L},\\
\frac{d}{dt}E_{mg}=-H_{0}\dot{M}_{z}+D_{x}M_{\mathrm{x}}\dot{M}_{x}\,.
\end{array}
\label{Eng}%
\end{equation}
We focus in the following on the regime $H_{0}\ll D_{x}M_{\text{s}}$, which
usually holds for thin films and not too strong fields, in which the
magnetization motion is elliptical with long axis in the plane and small
$M_{x}$ even for larger precession cones. Disregarding terms containing higher
powers of $M_{x}$ and averaging over one period as indicated by $\left\langle
...\right\rangle $%
\begin{equation}%
\begin{array}
[c]{c}%
\left\langle \frac{dE_{me}}{dt}\right\rangle +\left\langle 2\beta
E_{me}\right\rangle =-VD_{x}\left\langle M_{z}M_{x}\dot{\varphi}\right\rangle
,\\
\left\langle \frac{dE_{mg}}{dt}\right\rangle +\left\langle \alpha D_{x}%
^{2}M_{\text{s}}M_{x}^{2}\right\rangle =D_{x}\left\langle M_{z}\dot{M}%
_{x}\varphi\right\rangle .
\end{array}
\label{Eng1}%
\end{equation}
By adiabatic shaping of time-dependent magnetic fields $H_{0}(t)$ we can keep
the two subsystems at resonance at all times ($H_{0}(t)$ does not change much
since we do not consider large angle cones that are very close to the
antiparallel configuration). The slow dynamics $\varphi(L,t)\sim
A(t)e^{i\left(  \omega+\pi/2\right)  t}$ and $M_{x}\sim W(t)e^{i\omega t}$ in
time domain is then governed by the equation:%
\begin{equation}%
\begin{array}
[c]{c}%
\dot{A}+\beta A=-g\omega_{m}/M_{\text{s}}(-1+\frac{D_{x}W^{2}}{4M_{\text{s}%
}H_{0}})W\,,\\
\dot{W}+\alpha^{\prime}\omega W=\omega M_{\text{s}}(-1+\frac{D_{x}W^{2}%
}{4M_{\text{s}}H_{0}})A\,,
\end{array}
\label{oscillations1}%
\end{equation}
we introduced a frequency $\omega_{m}=\gamma\sqrt{D_{x}M_{\text{s}}H_{0}(t)}$
that at $t=0$ coincides with the frequency of fast oscillations $\omega$.
Substitutions $\varphi(L,t)\sim\widetilde{A}(t)e^{i\omega t}$ and $M_{x}%
\sim\widetilde{W}(t)e^{i(\omega-\pi/2)t}$ corresponding to $\pi/2$-shifted
harmonics are also solutions, and the initial conditions determine the linear
combination of two envelope functions, \textit{i.e}. the beating pattern of
two hybridized polariton modes \cite{Kovalev:apl03}. When initially all energy
is stored in one degree of freedom, $M_{x}$ is $\pi/2$ shifted from
$\varphi(L,t)$ and $A_{2}(t)=0$. Eqs. (\ref{oscillations1}) describe a
(damped) harmonic oscillator with frequency $\sqrt{g\bar{\omega}_{m}\omega}%
\ll\omega$ when $D_{x}W/4\ll MH_{0}$ (since $\omega_{m}$ does not change much
we replaced it by averaged $\bar{\omega}_{m}$). Such oscillatory behavior
persists for general angles (except for motion with very large angle cones
close to the antiparallel configuration). This is illustrated by Fig.
\ref{fig10} which shows a numerical simulation of Eq. (\ref{MME}) for an
undamped system excited at $t=0$ by a magnetic field $\mathbf{H}_{0}$ at an
angle $2\pi/3$ with the initial magnetization. The number of periods necessary
to transfer all energy from one subsystem to the other is therefore given by
$\sim1/\left(  4\sqrt{g}\right)  $. Eq. (\ref{oscillations1}) also shows that
for damping constants $\alpha>\varphi_{0}/\pi$ or $\beta/\omega>\sqrt{g}/\pi$
the beating is suppressed.

Fig. \ref{fig10} illustrates that the mechanical system absorbs energy from
the magnetic subsystem and gives it back repeatedly in terms of violent
oscillations that are modulated by an envelope function on the time scale
derived above in Eq. (\ref{oscillations1}) as plotted by dotted line. When the
envelope function vanishes the magnetization is reversed and the systems seems
to be at rest. However, since the energy is not dissipated, the momentarily
silence is deceptive, and the beating pattern repeats. An efficient coupling
requires that the frequencies of the subsystems are close to each other at
each configuration, which was achieved in the simulation by the adiabatic
modulation of the magnetic field $H_{0}$ according to Fig. 2. However, the
reversal process is robust; an estimate from Eqs. (\ref{Eng1}) for the
necessary proximity of the resonant frequencies of the mechanical and the
magnetic subsystems is $\Delta\omega\sim\sqrt{g}\omega$. In that case, the
above estimates still hold.

Let us now consider the magnetization reversal by an antiparallel magnetic
field $H_{0}(t)$ that undergo slow change of amplitude to keep the subsystems
at resonance. This method of magnetization reversal by oscillating
demagnetizing field due to mechanical oscillations is strongly analogous to
the reversal by time-dependent magnetic fields. \cite{Sun:2005} However, in
contrast to the time-dependent field reversal, we vary the field $H_{0}(t)$
but not the rf fields so avoiding complicated time dependence of rf fields. \cite{Sun:2005} 
We can calculate the dependence $H_{0}(t)$ in advance
for a specific sample, or we can create feedback circuit by connecting
metallic contacts to the magnetic film, thus monitoring the dynamics. Since we
assume zero temperature, the dynamics in Fig. \ref{fig11} is initiated by
assuming a tiny angle of magnetization at $t=0$. We wish to illustrate here
that making use of the magnetoelastic coupling can accelerate the reversal
importantly. We can suppress the backflow of mechanical energy for example by
a sufficiently damped mechanical subsystem, as evident in Fig. \ref{fig6} by
comparing the two curves for $\beta\sim0$ and $\beta\sim0.02$ with vanishing
Gilbert damping, $\alpha=0$. Alternatively, we may detune the external
magnetic field out of the resonance precisely after the first reversal,
effectively rectifying the energy flow from the magnetic into the mechanical
subsystem. We observe that even without any intrinsic damping ($\alpha
=\beta=0$) the unwanted \textquotedblleft ringing\textquotedblright\ can be
strongly suppressed (dashed line in Fig. \ref{fig6}).

The experimental realization of such magnetization reversal will be a
challenge since the cantilever has to preferably work at high frequencies
$\sim$ 1 GHz. In the resonant reversal, a significant coupling strength of
$g\sim10^{-3}$ requires that one tenths of the cantilever volume is a
ferromagnet. We investigated here the non-linear dynamics of coupled magnetic
and mechanical fields for a cantilever with a ferromagnetic tip. Employing the
new dissipation channels, we propose new strategies for fast magnetization
reversal and suppressed \textquotedblleft ringing\textquotedblright.  We can
make use of the additional mechanical damping or shape the external magnetic
field pulses, thus quickly channeling-off magnetic energy when damping is weak.

\subsection{Non-resonant magnetization oscillations and reversal}

We propose a non-resonant mechanical reversal scheme analogous to
\textquotedblleft precessional\textquotedblright\ switching
\cite{Gerrits:nat02}. The effective field $\mathbf{H}_{\mathrm{eff}}$ (see Eq.
(\ref{MME})) has a component perpendicular to the plane of the film $H_{x}\sim
M\varphi$. Under a sudden mechanical twist this component acts like a
transverse magnetic pulse about which the precessing develops. Alternatively,
we can suddenly release a twist preinstalled on the cantilever. This could be
achieved by an STM tip bonded to an edge of the cantilever at ($L$, $d$) and
pulling it up slowly to the breaking point. The mechanical response should be
fast, \textit{i.e.} react on a time scale $(\gamma\varphi\nu M)^{-1},$ but
there are no resonance restrictions now. We integrate the equation of motion
numerically for a strongly damped cantilever $\beta/\omega\sim0.15$ initially
twisted by $\varphi=0.2$ and suddenly released at $t=0$. We reintroduce an
easy axis anisotropy described by $DM_{z}\mathbf{z}$. Adopt $D=0.05$,
$\alpha=0.01$ and no external fields. Fig. \ref{fig12} displays the desired
reversal. The rather severe overshoot, as in the case of the precessional
switching technique, can be minimized by carefully engineering the mechanical
actuation to be closer to the optimum \textquotedblleft
ballistic\textquotedblleft\ path between $M_{z}=\pm1$.

Summarizing, we demonstrate here a precessional reversal scheme based on the
mechanically generated out-of-plane demagnetizing field without applied
magnetic fields. In this reversal scheme, the sharp control of the resonance
condition is not required, however, the cantilever has to be fast determining
the reversal time.

\section{Conclusion}

This paper reports a detailed study of magnetomechanical effects in small
magnetic cantilevers, summarizing some old
results\cite{Kovalev:apl03,Kovalev:prl05} and making several generalizations.
We proved that for small magnetization oscillations close to the mechanical
resonance frequency, a ferromagnetic cantilever behaves like the dielectric
cantilever with a small magnet at the tip analyzed before. Such a
ferromagnetic cantilever has an enhanced coupling of mechanical and magnetic
degrees of freedom and thus is more suitable for observation of
magnetomechanical effects. A metallic ferromagnetic cantilever can be
integrated into magnetoelectronic circuits as shown in Sec. IV. Such an
integrated device has the potential as a fast transducer of mechanical motion
as an alternative to previous designs that take advantage of magnetomotive
forces. In our case, the magnetomotive forces are supplanted by spin-transfer
torques. Our strategy; that is to say to avoid magnetic fields, is similar to
what happens in the field of random access memories in which there is a
considerable effort to replace magnetic fields by employing spin-trasfer
torques. The technique presented in this paper strongly relies on resonant
magnetovibrational coupling in which the magnetic torques can be effectively
transformed into mechanical torques and \emph{vice versa}. In order to see
magnetomechanical torques in experiement, one needs to be able to handle small
structures on micro and nano scale, the magnets have to be small enough to
form a single domain. We beleive that the integration of magnetoelectronics
and magnetomechanics is possible and should lead to devices with new functionalities.

\section*{Acknowledgment}
We thank Yaroslav Tserkovnyak for helpful discussions. This work has been supported by the Dutch FOM Foundation and the Research Council of Norway.

\newpage
\begin{figure}[ptb]
\caption{A magnetomechanical
cantilever supporting magneto-vibrational modes. On a dielectric substrate
(such as Si) a single-domain ferromagnetic film is deposited at the free end.}%
\label{fig1}%
\end{figure}

\begin{figure}[ptb]
\caption{Dependence of FMR
broadening on aspect ratio $m=a/c$ for ellipsoid with the semi-axes $a$ and
$b=c$ ($H_{0}=M_{\text{s}}$).}%
\label{fig2}%
\end{figure}

\begin{figure}[ptb]
\caption{The period
$T_{1}$ ($E_{mg}<1$) and $T_{2}$ ($E_{mg}>1$) of the magnetization dynamics in
Eqs. (\ref{T1/2}) in units $2\pi/\left(  \gamma\sqrt{(H_{0}+D_{x}M)H_{0}%
}\right)  $ as a function of energy in units $H_{0}M$. The inset shows a plot
of typical trajectories at different energies on the unit sphere ($D_{x}M=10H_{0}$}).%
\label{fig3}%
\end{figure}

\begin{figure}[ptb]
\caption{Dependence of the
resonance frequency of the coupled motion on the FMR frequency $\omega
_{\text{m}}$ of the uncoupled magnetization ( $g\sim0.001$).}%
\label{fig4}%
\end{figure}

\begin{figure}[ptb]
\caption{The oscillator strength
corresponding to each resonance in Fig. 4 in arbitrary units plotted by full
lines and their widths plotted by dashed lines ($g\sim0.001$, $\beta/\omega>\alpha'
$).}%
\label{fig5}%
\end{figure}

\begin{figure}[ptb]
\caption{Amplitude of mechanical
oscillations along one of the resonances in Fig. \ref{fig4}. The maximal
efficiency of the \textquotedblleft nanomotor\char`\"{} occurs when both
subsystems are at resonance.}%
\label{fig6}%
\end{figure}

\begin{figure}[ptb]
\caption{Electro-magneto-mechanical
device.}%
\label{fig7}%
\end{figure}

\begin{figure}[ptb]
\caption{Dependence of the normalized
impedance on the AC current frequency ($\alpha^{\prime}=0.02$).}%
\label{fig8}%
\end{figure}

\begin{figure}[ptb]
\caption{Dependence of the normalized
impedance on the AC current frequency ($\omega_{m}=\omega_{e}$, $\alpha^{\prime}=0.02$, $\beta
/\omega=0.002$).}%
\label{fig9}%
\end{figure}

\begin{figure}[ptb]
\caption{Time-dependent response of
the magnetomechanical system to an external magnetic field switched on at
$t=0$ at an angle $2\pi/3$ with the initial magnetization, in the absence of
dissipation. Plotted are the $\varphi(L)$ and $z-$components of the
magnetization ($D_{x} M=10H_{0}$).}%
\label{fig10}%
\end{figure}

\begin{figure}[ptb]
\caption{Time-dependent response
of the magnetomechanical system to an external magnetic field switched on at
$t=0$ ($D_{x} M=10H_{0}$, $\alpha=0$). The importance of mechanical damping can
be seen by comparing the dotted line ($\beta=0$) with the full line
($\beta=0.02$). The dashed line illustrates the dynamics when the subsystems
brought out of resonance by reducing the external magnetic field to half of
its initial value at $t=6.5$ ($\beta=0$).}%
\label{fig11}%
\end{figure}

\begin{figure}[ptb]
\caption{Magnetization switching
without external field by an initially twisted cantilever that is released at
$t=0$. The inset shows the corresponding magnetization trajectory on the unit
sphere.}%
\label{fig12}%
\end{figure}

\newpage

\newpage
\newpage

\centerline{\includegraphics{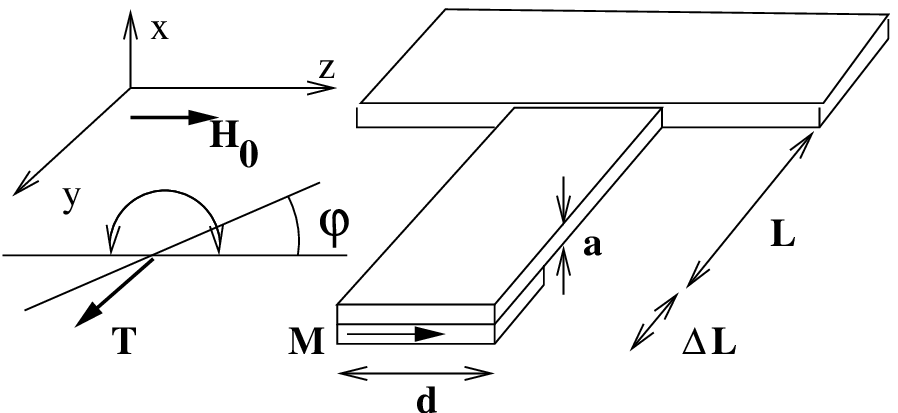} }
\vspace{1cm}
\centerline{\Large{Fig. 1}}

\newpage
\centerline{\includegraphics{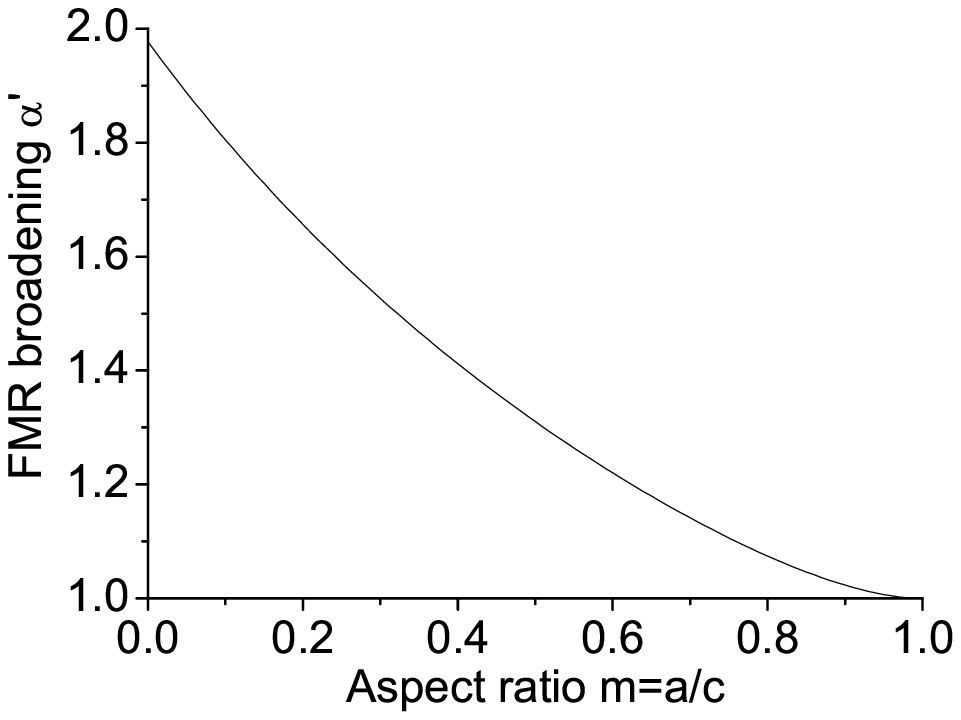} }
\vspace{1cm}
\centerline{\Large{Fig. 2}}

\newpage
\centerline{\includegraphics[  scale=0.5]{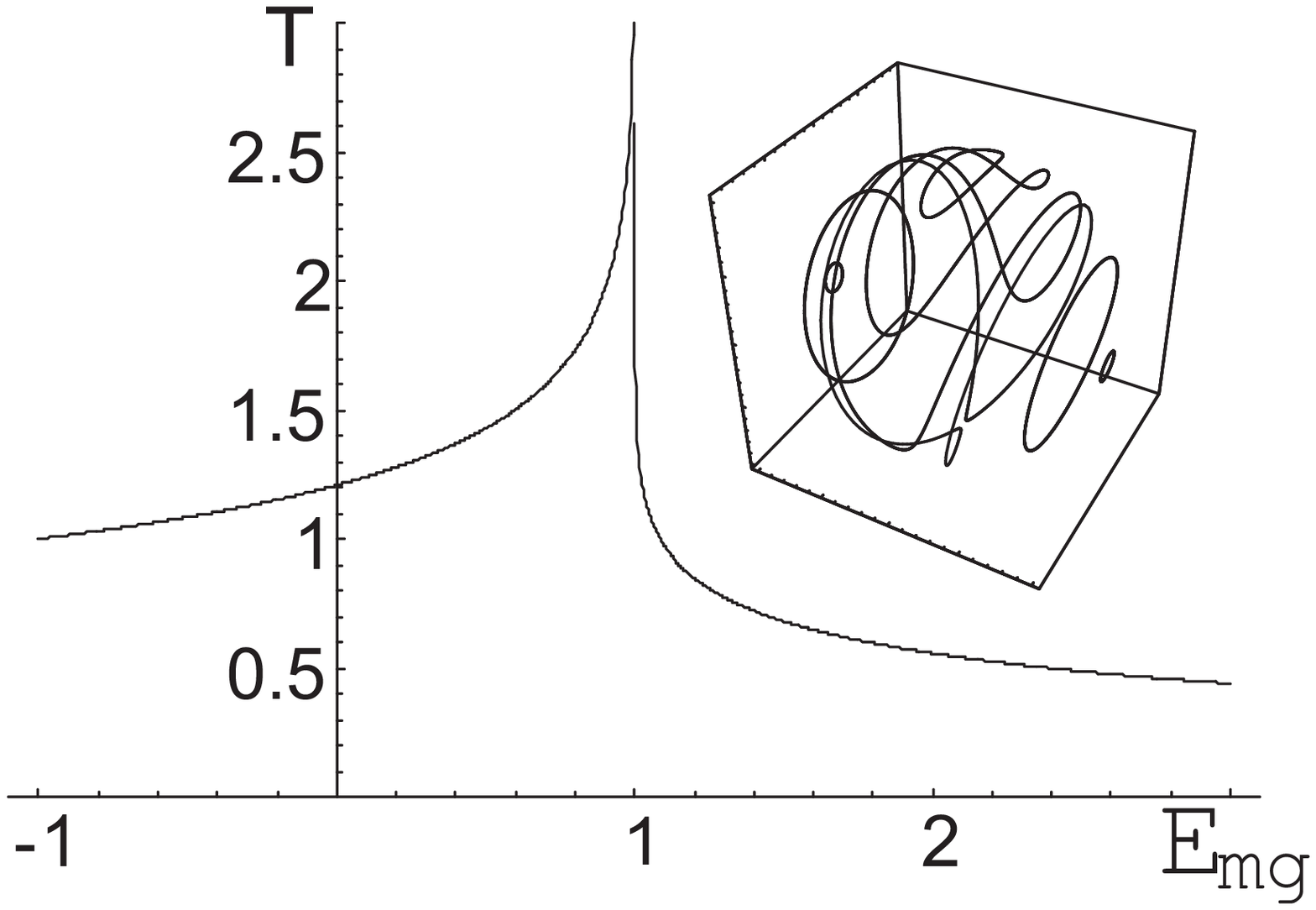} }
\vspace{1cm}
\centerline{\Large{Fig. 3}}

\newpage
\centerline{\includegraphics{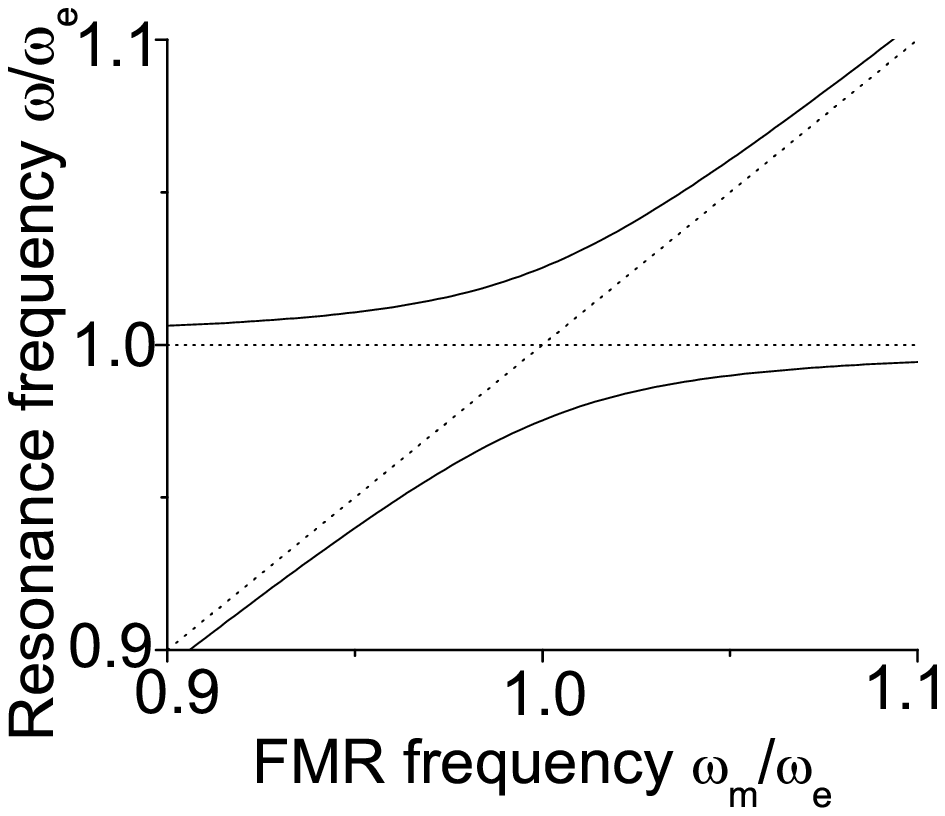} }
\vspace{1cm}
\centerline{\Large{Fig. 4}}

\newpage
\centerline{\includegraphics{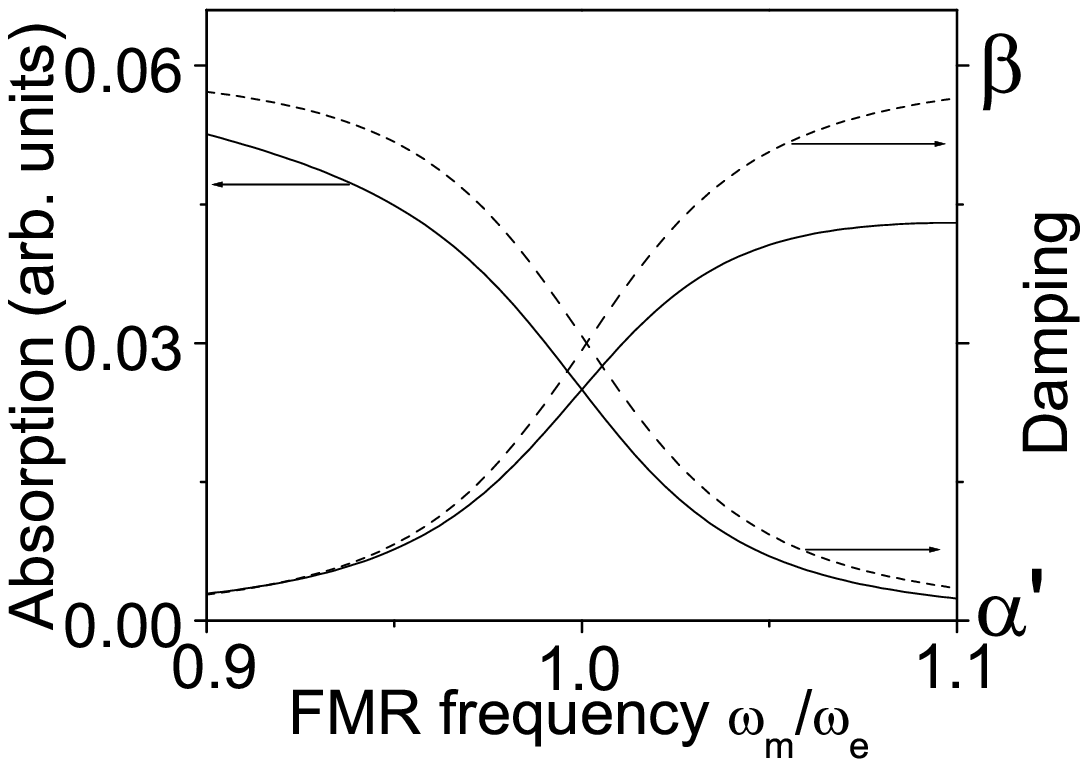} }
\vspace{1cm}
\centerline{\Large{Fig. 5}}

\newpage
\centerline{\includegraphics{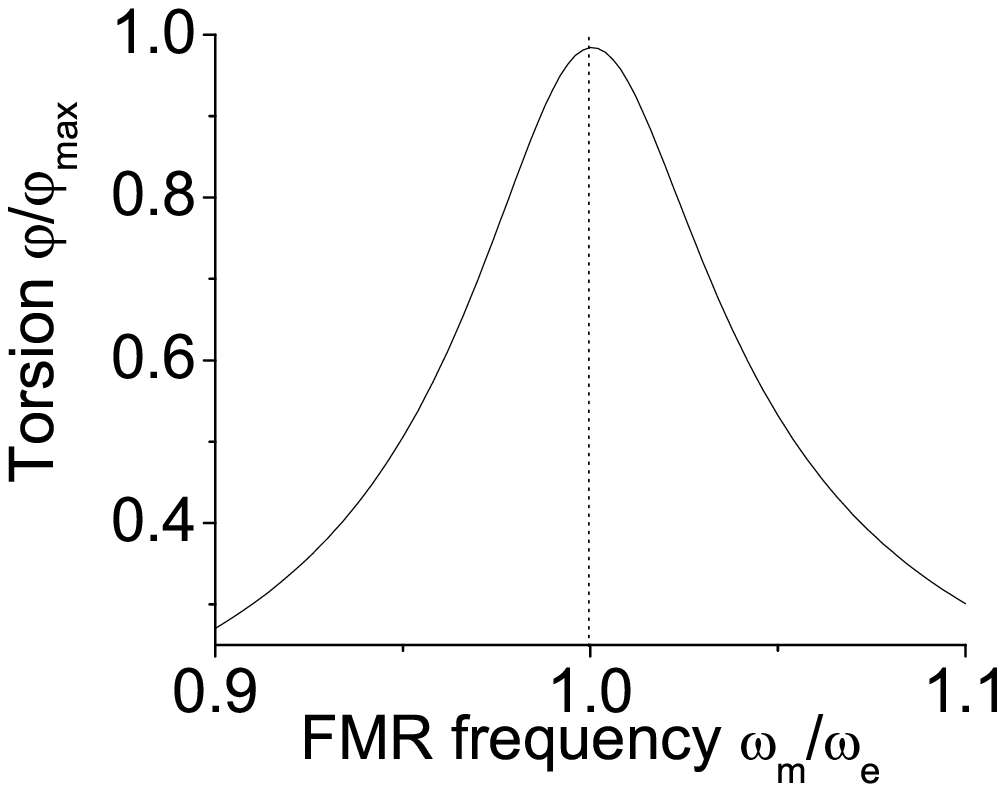} }
\vspace{1cm}
\centerline{\Large{Fig. 6}}

\newpage
\centerline{\includegraphics{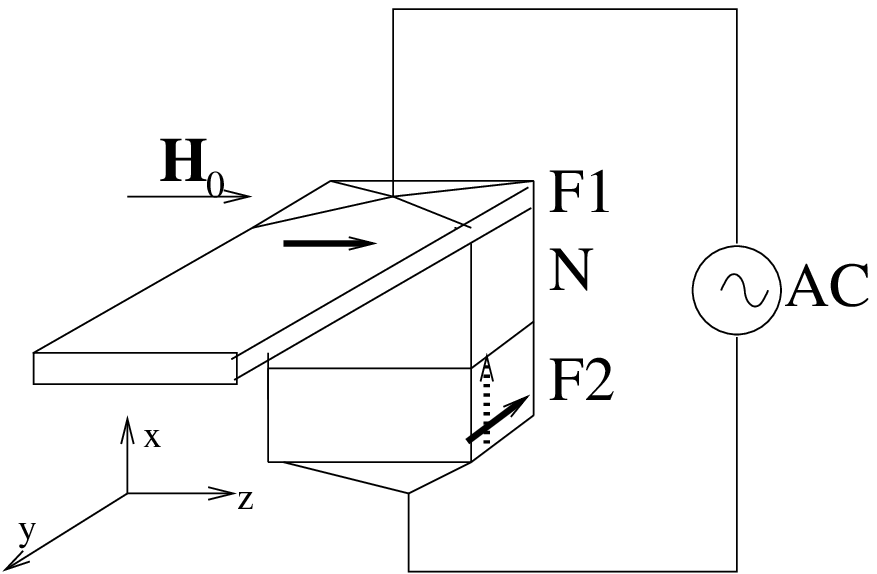} }
\vspace{1cm}
\centerline{\Large{Fig. 7}}

\newpage
\centerline{\includegraphics{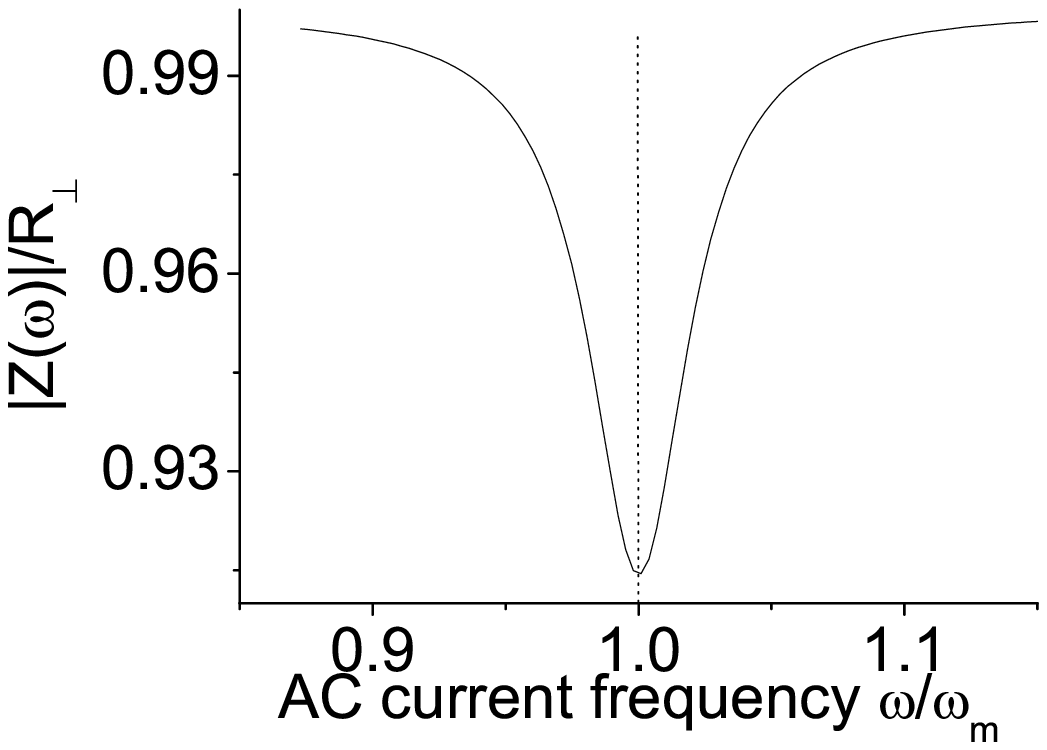} }
\vspace{1cm}
\centerline{\Large{Fig. 8}}

\newpage
\centerline{\includegraphics{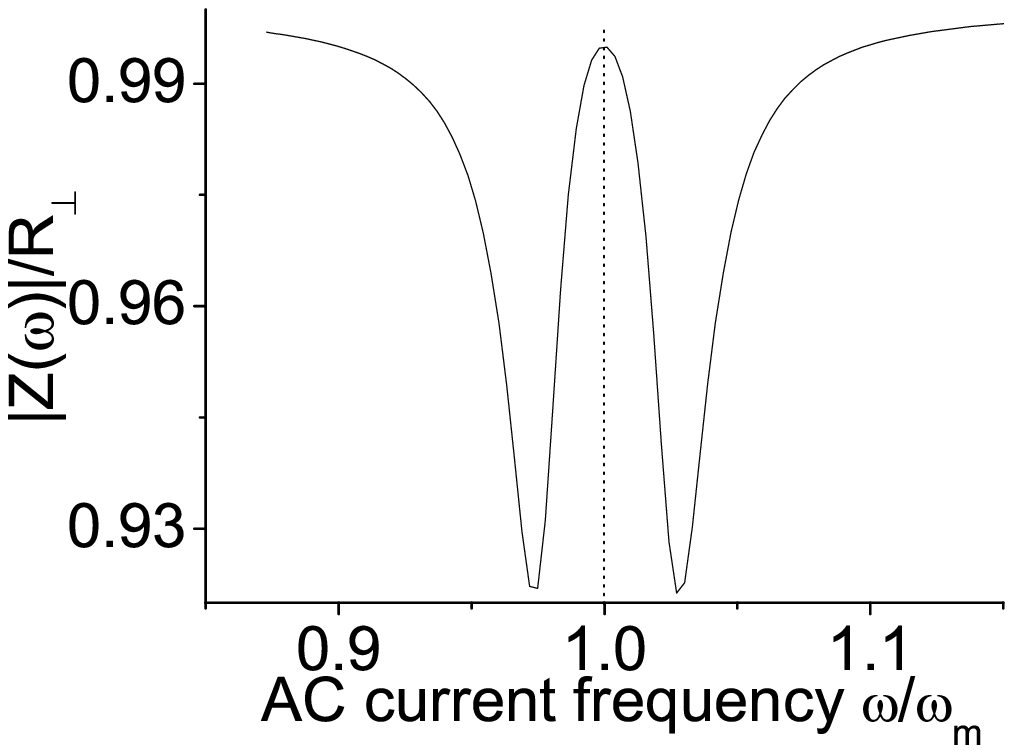} }
\vspace{1cm}
\centerline{\Large{Fig. 9}}

\newpage
\centerline{\includegraphics{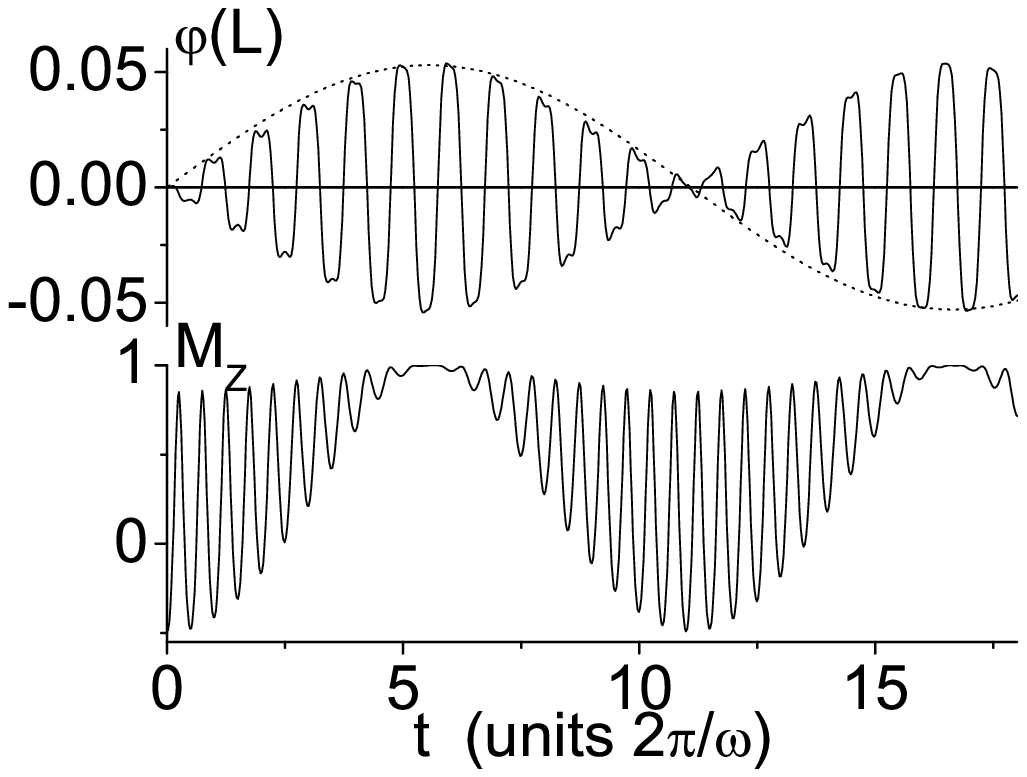} }
\vspace{1cm}
\centerline{\Large{Fig. 10}}

\newpage
\centerline{\includegraphics{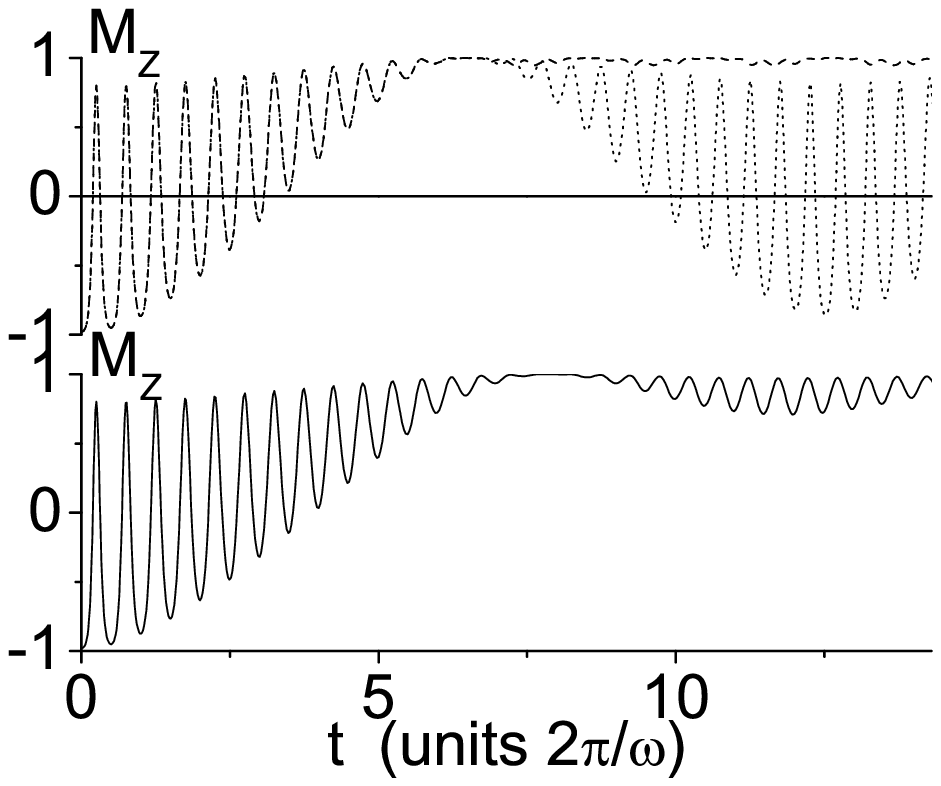} }
\vspace{1cm}
\centerline{\Large{Fig. 11}}

\newpage
\centerline{\includegraphics{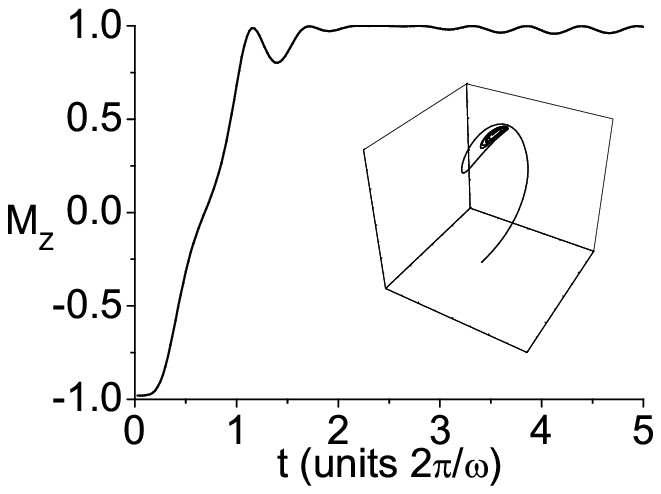} }
\vspace{1cm}
\centerline{\Large{Fig. 12}}

\end{document}